\documentclass[10pt,letterpaper]{article}
\usepackage[top=0.85in,left=2.75in,footskip=0.75in]{geometry}

\usepackage{changepage}

\usepackage[utf8]{inputenc}

\usepackage{textcomp,marvosym}

\usepackage{fixltx2e}

\usepackage{amsmath,amssymb}

\usepackage{cite}

\usepackage{nameref,hyperref}

\usepackage[right]{lineno}

\usepackage{microtype}
\DisableLigatures[f]{encoding = *, family = * }

\usepackage{rotating}

\usepackage{ifthen}
\newcounter{submission}
\ifthenelse{\value{submission} > 0}{
\usepackage{setspace} 
\doublespacing
}{ }

\usepackage{here}
\usepackage{docmute}

\raggedright
\usepackage[aboveskip=1pt,labelfont=bf,labelsep=period,justification=raggedright,singlelinecheck=off]{caption}

\makeatletter
\renewcommand{\@biblabel}[1]{\quad#1.}
\makeatother

\date{}

\usepackage{lastpage,fancyhdr,graphicx}
\usepackage{epstopdf}

\begin{document}
\vspace*{0.35in}

\begin{flushleft}
{\Large
\textbf\newline{Sam2bam: High-Performance Framework for NGS Data Preprocessing Tools}
}
\newline
\\
Takeshi Ogasawara\textsuperscript{1*},
Yinhe Cheng\textsuperscript{2},
Tzy-Hwa Kathy Tzeng\textsuperscript{3}

\bigskip
\bf{1} IBM Research -- Tokyo, Tokyo, Japan
\\
\bf{2} IBM Systems, Austin, TX, USA
\\
\bf{3} IBM Systems, Poughkeepsie, NY. USA
\\
\bigskip

* takeshi@jp.ibm.com

\end{flushleft}
\section*{Abstract}
This paper introduces a high-throughput software tool framework called {\it sam2bam} that enables  users to significantly speedup pre-processing for next-generation sequencing data. The sam2bam is especially efficient on  single-node multi-core large-memory systems.
It can reduce the runtime of data pre-processing in marking duplicate reads  on a single node system by 
156-186x 					
 compared with de facto standard tools. 
The sam2bam consists of parallel software components that can fully utilize the multiple processors, available memory, high-bandwidth of storage, and hardware compression accelerators if available.

The sam2bam provides file format conversion between well-known genome file formats, from SAM to BAM, as a basic feature.  Additional features such as analyzing, filtering, and converting the input data are provided by {\it plug-in} tools, e.g., duplicate marking, which can be attached to sam2bam at runtime. 

We demonstrated that sam2bam could significantly reduce the runtime of NGS data pre-processing from about two hours to about one minute for a whole-exome data set on a 16-core single-node system using up to 130 GB of memory.
The sam2bam could reduce the runtime for whole-genome sequencing data from about 20 hours to about nine minutes on the same system using up to 711 GB of memory.

\section*{Introduction}
The rapid advance of sequencing technology and its falling cost are driving the use of next-generation sequencing (NGS) in a great variety of domains.
The volume of data generated by NGS was projected to double every five months \cite{data_overload}. Processing raw data from a sequencer and translating it into usable insights in the timely manner is a formidable task.
Typical 50x coverage of whole genome sequencing (WGS)
can easily generate up to ~500 GB FASTQ files. 
The data processes on modern computers is still very time-consuming for such huge data sets.
For typical 50x coverage of WGS data, the processing time for the Broad's Genome Analysis Toolkit (GATK) Best Practices pipeline \cite{gatkbest}  from reference alignment to variant calling can take up to a day or days to finish \cite{intelGATK}. 

Genome data analysis pipelines involve the data pre-processing steps before variant calling, which are necessary to achieve accurate variant calling. They scan the input data, analyze reads, and filter out  reads that can affect  accuracy. 
The pre-processing steps take FASTQ-format \cite{fastq} files
as the input and produces a compressed binary file in the BAM format \cite{samtools}, which  are widely accepted as  common file formats to represent aligned sequenced data.
The calibrated BAM file is then used in the variant discovery to identify the sites where the data displays variation relative to the reference genome.

The pre-processing steps for SAM parsing, sorting, duplicate marking, and BAM file compression can take tens of hours 
for a WGS SAM file. Their total runtime dominates the whole pre-processing workflow (explained in the next sub-section)
and is a clear performance bottleneck. The purpose of sam2bam is to improve the efficiency of this pre-processing through fully utilizing available CPUs and memory.

The overall architecture of the tools should be redesigned so that the computer resources are fully utilized to significantly reduce the runtime of such data pre-processing steps  (e.g., by 100x). Current major software tools are single- or partially multi-threaded. 
Partially multi-threaded tools 
usually have to wait for data generated from single-threaded components because every component (single- or multi-threaded) is executed one by one. 
Therefore, they can not fully utilize multiple CPUs that are available at all times. 
Single-threaded components become bottlenecks in  performance on multi-CPU systems. For example, suppose that 80\% of the runtime is executed by multi-threads and the remaining time is executed by a single thread. Speed-up for such tools by using multiple CPUs is limited to 5x even if hundreds of CPUs are available on the system.

The sam2bam simultaneously executes functional components (e.g., file I/O, SAM parsing, and data compression) to achieve further speed-up on such a many-CPU system, instead of  executing the components one by one in a big loop. 
The components are combined as a pipeline. In addition, most of the steps are multi-threaded. An appropriate number of CPUs are allocated to each component so that no components become a bottleneck in the pipeline. 
The sam2bam can achieve more than 100x speed-up on a single node system with these redesigned framework for NGS data pre-processing.

\subsection*{Pre-processing Steps for Variant Discovery}
To maximize the accuracy of variant discovery, pre-processing steps are necessary to prepare the data for analysis. Pre-procesing is also recommanded in GATK Best Practices \cite{gatk-preprocess}. Pre-processing starts with FASTQ-format files and ends in a calibrated BAM file.
For the DNA data, pre-processing usually involves the following steps.
\begin{enumerate}
\item{\bf Mapping the sequence reads to the reference genome} Usually this step is done by BWA mem \cite{bwa} or other reference alighment tools. SAM files are generated.
\item{\bf Sorting the sequence reads based on coordinates} This step is usually done by Picard SortSam or samtools sort. Some tools that are used in the following steps such as Picard MarkDuplicates require the sorted input files. 
\item{\bf Marking duplicate alignments} This step is commonly done by Picard MarkDuplicates tool to remove the alignments of duplicate reads.
\item{\bf Performing local realignment around indels} This is usually done by GATK RealignerTargetCreator and IndelRealigner tools to reduce artifacts produced in the regions around the indels. 
\item{\bf Recalibrating the base quality score} This step is usually done by GATK BaseRecalibrator and PrintReads tools to improve the accuracy of base quality scores which the variant calling step relies on.
\end{enumerate}

Preprocessing is very time cosuming usually tens of hours for a WGS dataset and hours for a WEX dataset. Among the five steps of preprocessing, the sorting and duplicate marking steps take most of the runtime, usually ranging from 60-70\% of the total pre-processing time, depending on the tools used and the test case size. Therefore the sorting and duplicate marking steps are identified as a bottleneck of overall pre-processing steps, and sam2bam is focused on improving the performance of these two steps by redesigning the framework, paralleling most of the process and taking advantage of hardware compression when available.

\section*{Design and Implementation}

\begin{figure}[H]
\caption{{\bf Architecture for sam2bam without analyzer plug-ins.} Pipeline is configured with  plug-in code that filters out  data. Gray boxes indicate steps in pipeline. Steps that have multiple boxes are multi-threaded. Blue boxes denote  files in storage. Light-blue boxes denote  data in memory. Light green boxes denote  plug-in code.}
\ifthenelse{\value{submission} > 0}{ }{
\includegraphics[trim={0cm 7.5cm 18cm 0cm},height=6.5cm,clip]{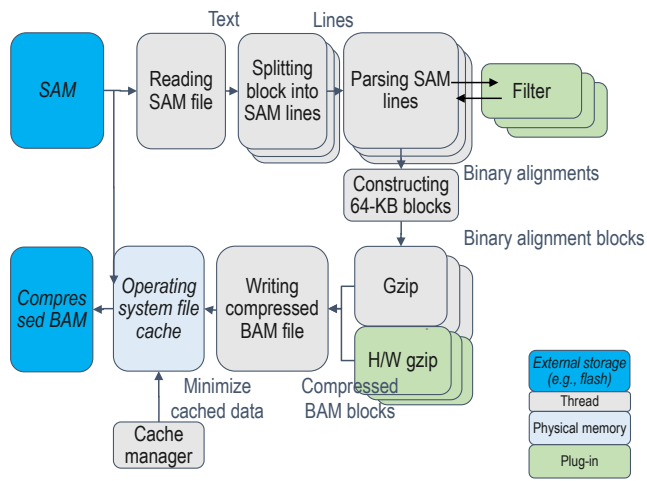}	
}
\label{arch-without-baminfo}
\end{figure}

\begin{figure}[H]
\caption{{\bf Architecture for sam2bam with analyzer plug-ins.} Alignment database is created when analyzer plug-ins are enabled. Binary alignments that are produced by  SAM parsing are placed in either  main memory or  external storage so that they can be used later for generating  compressed BAM file by using second half of  pipeline. Alignment database has  summarized information on binary alignments that is used by  analyzer plug-ins. }
\ifthenelse{\value{submission} > 0}{ }{
\includegraphics[trim={0cm 4cm 10cm 0cm},width=14cm,clip]{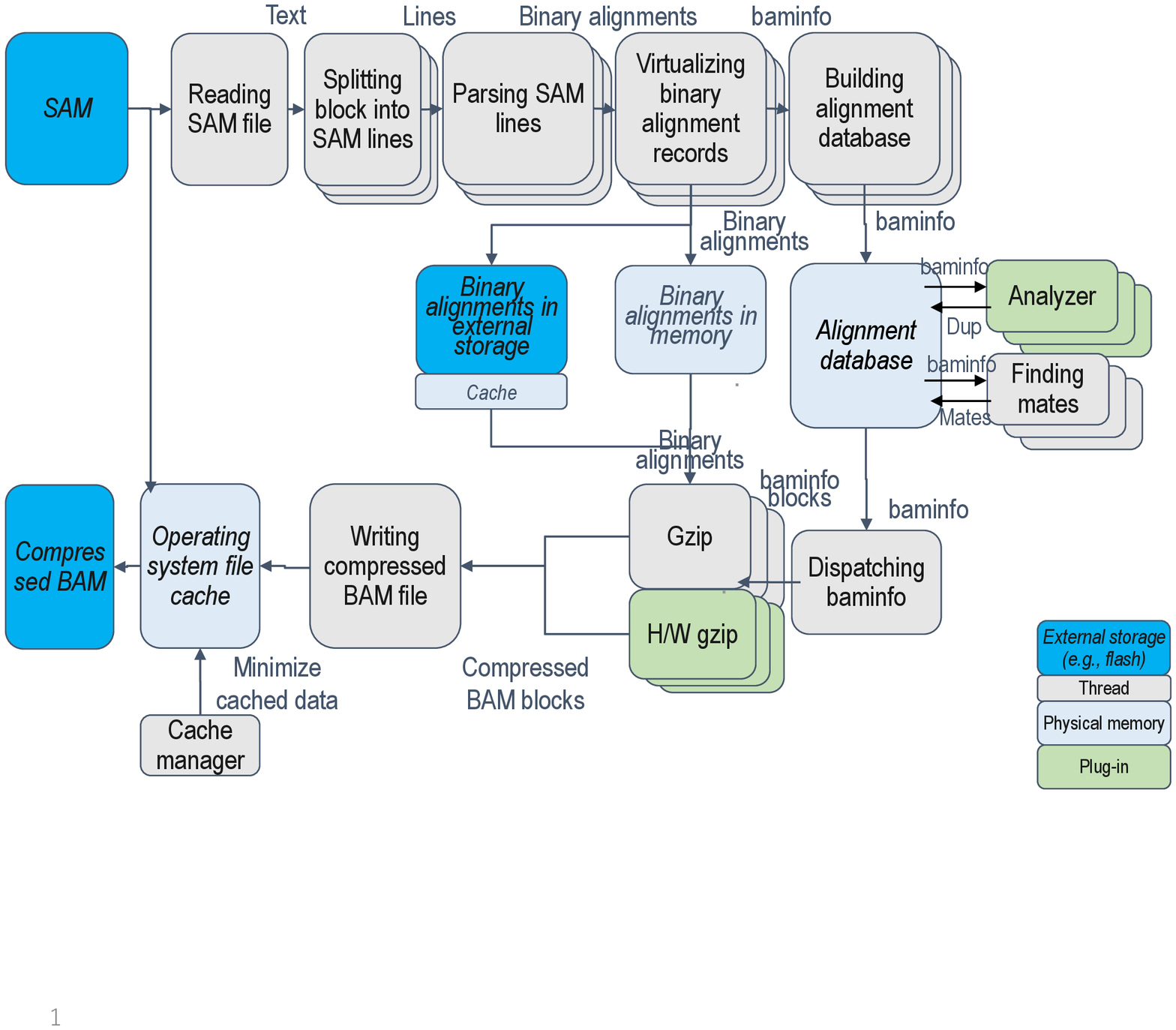}	
}
\label{arch-with-baminfo}
\end{figure}

The design goal of sam2bam was to provide a high-throughput framework to process genome files at a rate of gigabytes per second (GB/s). The framework consists of a data pipeline that converts the data format from SAM to BAM as outlined in Figs 
\ref{arch-without-baminfo} and \ref{arch-with-baminfo}. Data format conversion is divided into multiple steps. Many of these steps are multi-threaded, while a few steps that order the data stream are single-threaded. More CPUs are allocated for more complex steps so that such steps are not bottlenecks in  performance. Each step continuously processes data by using  CPUs as long as the data drives from the previous step. 
While sam2bam provides a data processing pipeline, it uses samtools/high-throughput sequencing library (HTSLIB) \cite{samtools} for the data structures and utility functions for handling the SAM and BAM data formats. 

\subsection*{Plug-in Codes}

{\it Plug-in codes} that analyze, filter, and modify data can be attached to sam2bam at runtime. We can develop the plug-in codes and run them on the high-throughput framework. There are three types of  plug-in codes. 
\begin{itemize}
\item{\it Filter } The filter plug-in code can be inserted as an additional step of the pipeline. It analyzes  input data and determines if the data meets the criteria that the plug-in has. For example, with a filter plug-in that only includes read alignments that overlap a given region of the reference genome, the produced BAM file only includes  alignments that overlap the specified region. 

\item{\it Accelerator } The accelerator plug-in code improves the target pipeline step by using  hardware accelerators, such as field-programmable gate arrays (FPGAs). Compression acceleration is currently supported in sam2bam. It produces a compressed BAM file using the standard compression library (or {\tt zlib}) \cite{sambam}. 
If sam2bam detects an accelerator, it automatically offloads  compression  to  hardware while carrying out compression with software. An accelerator that provides the same application programming interface (API) as {\tt zlib} can be enabled by using the accelerator plug-in.

\item{\it Analyzer } The analyzer plug-in code analyzes a set of  read alignments and modifies them. If the analyzer plug-in code is attached to sam2bam, the latter runs the first half of the pipeline that parses  data in the SAM format and pools the alignment information in the system. How the alignment information is pooled will be explained in the next subsection (\nameref{virtualizing}: items \ref{virtualizing} and \ref{build_alignment_db}). When all alignment information is pooled in the system, the analyzer plug-in scans the pooled information, analyzes it, and generates  output on the basis of the analysis. When the analyzer plug-in has completed analysis, the pooled data are transferred to the second half of the pipeline to produce a compressed BAM file. For example, with an analyzer plug-in for marking duplicate alignments, the alignments that were mapped to the same reference region are identified and are marked as duplicates except for the one that has the highest level of quality.
\end{itemize}

\subsection*{Pipeline Configuration}

The entire process of sam2bam conversion is split into functional stages such as file reading, SAM line detection, and SAM parsing. The stages run in parallel. Also each stage can process multiple data blocks in parallel. 
\begin{enumerate}

\item\label{sam_reading}{\bf Reading a SAM file} Data blocks (e.g., 64 KB each) are read from a SAM file, which is a sequence of text lines (called SAM lines). 
Data blocks do not always end at the boundaries between SAM lines. 
This step adjusts the boundaries between the blocks so that the new blocks end at the SAM line boundaries by scanning  each block from the end to the beginning until the first new-line character is found to enable multi-threads to process the blocks independently on the next step.

\item\label{line_splitting}{\bf Splitting a block into SAM lines} SAM lines are extracted from the data blocks transferred from the previous stage. The data blocks are scanned from the beginning to the end to find new-line characters, which separate SAM lines. A major effort in this stage is to find these characters by using a function of the standard C-language library ({\tt memchr}). The performance of the function is optimized by using vector instructions, if they are available \cite{memchr}. A scanned data block is transferred to the next pipeline stage, which parses SAM lines with the positions where SAM lines start.

\item\label{sam_parsing}{\bf Parsing SAM lines} Binary alignment records are created from the SAM lines. The sam2bam locates a SAM line in the data block at each line position calculated by the previous stage, parses the SAM line by using a library function of Samtools ({\tt sam\_parse1}), and creates a binary alignment record in the data format  used in Samtools ({\tt bam1\_t}). The binary alignment record contains the same contents as the corresponding SAM/BAM alignment. Each binary record has a global sequence number, which is created  on the basis of the block number and a local number of the record within the block. This global sequence number is used when analysis tools need to know which record has appeared first in the input file for a given set of records.

\item\label{virtualizing}{\bf Virtualizing binary alignment records} The sam2bam supports two modes: all  the binary alignment records are placed in the main memory ({\it  memory mode}) or in external storage ({\it  storage mode}). If there is sufficient memory, the memory mode is recommended to be run to obtain the best performance; 
otherwise, the storage mode can be used.
The binary alignment records are moved to either the main memory or the external storage in this step. The binary alignment records are managed in the later stages in the virtual space of sam2bam. A virtual address is assigned to each binary alignment record. A virtual address can be translated into either a main memory address or an offset of the file in external storage. 

If sam2bam is  only invoked with plug-ins that access  streaming input data but not  pooled information, the input data are transferred from this step to the last step \ref{write_compressed_BAM_file}. In that case, only the memory mode is enabled. Also, steps \ref{build_alignment_db} and \ref{analyze_baminfo} are skipped. 

\item\label{build_alignment_db}{\bf Building  alignment database} 
Some plug-ins can start analysis after all  the input data are read. An alignment database is built for such plug-ins to provide  pooled information on the alignments in this step. Each database entry is represented by a data record called {\it baminfo}, which is unique to sam2bam. Baminfo is created from each binary alignment record but it only contains  information that can be used by the analyzer plug-ins in step  \ref{analyze_baminfo}.
It does not include  long string data, including reference sequence names, concise idiosyncratic gapped alignment report (CIGAR), sequences, base qualities, or optional fields. 
If any analyzer plug-in needs any of the omitted data in  step \ref{analyze_baminfo},
the plug-in summarizes the data and saves them into baminfo when creating baminfo.

The baminfo records are arranged in the reference genome position space to construct the alignment database. The baminfo records can be looked up from the alignment database by using the mapping positions of clipped sequences. The baminfo records that have the same mapping position are gathered and are found in the database by a single lookup. If there is an analysis that needs the functionality of looking up the baminfo records by using an unclipped mapping position, the baminfo records are arranged by using the unclipped position as well as the clipped position.

\item\label{analyze_baminfo}{\bf Analyzing baminfo records}
The plug-in code performs any analysis by using  information in the alignment database. Duplicate marking is an example of such code. Multi-threading can accelerate this step by decomposing the input space into sub spaces. For example, we can split the data set of the alignment database into $N*B$ blocks and allocate $B$ blocks to each of the $N$ threads.

\item\label{write_compressed_BAM_file}{\bf Writing a compressed BAM file}
This step groups the binary alignment records into ~64-KB blocks (BAM blocks), compresses them, and writes a sequence of the compressed BAM blocks to a BAM file. 
This step produces a sorted BAM file if the alignment database is available. The sam2bam does not require a separate step for sorting. If some plug-ins set up the alignment database, this step obtains  input by scanning the alignment database in the order of  sort keys and writes the alignments to the BAM file in the sort key order. Therefore, all the output from sam2bam is automatically sorted.
In contrast, if no analyzer plug-ins are used, the output is not sorted.

There is a dispatcher that splits a sequence of the binary alignment records into 64-KB sub-sequences by only using  their virtual addresses and their sizes. The dispatcher obtains  input data by traversing the alignment database or receiving them from  step \ref{virtualizing}. The dispatcher is single-threaded to ensure the order of the binary alignment records in each BAM block and the order of BAM blocks in the output. 
It does not construct actual BAM blocks by copying the data to avoid performance bottlenecks in the 
dispatcher. It instead  transfers each set of virtual addresses that construct a single BAM block to a thread pool for compression.

If the dispatcher receives the input data from  step \ref{virtualizing}, the binary alignment records in the physical memory provide their record sizes. Otherwise, the dispatcher traverses baminfo records in the alignment database to obtain the virtual addresses and  record sizes for the binary alignment records.

Multi-threads are used for compressing BAM blocks to accelerate this step. BAM blocks are constructed and compressed in parallel. The compression method is gzip, which is widely used in the real world \cite{samtools}.  If hardware compression accelerators are available in the system, sam2bam dynamically loads their library codes that can be called via the standard zlib API and it can offload compression to the accelerators.

\end{enumerate}

\section*{Results and Discussion}

To demonstrate that sam2bam can significantly reduce the runtime of marking duplicate alignments, we compared the runtime of sam2bam versus Picard \cite{picard}, which is a widely-used tool set that is also recommended in GATK best practices \cite{gatkbest-markdup}.

\subsection*{Experimental Environment}

\subsubsection*{Benchmark Data Sets}

Two data sets in the SAM format were used to evaluate performance. The first was a150x coverage of whole exome (WEX) data, and second was a 50x coverage of whole genome sequencing (WGS) data. 
The WEX data were part of the 1000 Genome Project data \cite{1000genome}. The input SAM file size was 52 GB.
The WGS data were part of  the Cancer Genome Atlas (TCGA) Benchmark 4 dataset, G15512.HCC1954.1 \cite{TCGA}. The input SAM file size was 546 GB.
The SAM files that we used were created by running Burrows-Wheeler Aligner (BWA) \cite{bwa} for the FASTQ-format data  converted from the original BAM files.

\subsubsection*{Sam2bam Configuration}

The sam2bam handled the SAM and BAM data formats by calling the modified code of samtools. The original code for samtools was obtained from its development repository as of August 2015. Two plug-ins were created for the performance evaluation: an analyzer plug-in for marking duplicate alignments and a compression accelerator plug-in. 
The source code of sam2bam and the instructions on how to build sam2bam are available from a GitHub repository (https://github.com/t-ogasawara/sam-to-bam).

The analyzer plug-in was enabled to mark duplicate alignments. The sam2bam created the alignment database, as was explained earlier in Section \nameref{build_alignment_db}: item \ref{build_alignment_db}.  The analyzer plug-in traverses the alignment database by using the unclipped position to find the candidates of duplicate alignments and finds  alignments that have the same beginning and end positions. It also finds their pairs by using  mate information that is available in the alignment database. The plug-in further uses the same criteria as Picard MarkDuplicates \cite{bamUtil} for the candidates to select one alignment among duplicates. The duplicates are analyzed in parallel by assigning the segmented unclipped position regions to threads.
This analyzer plug-in is provided as a pre-built library for POWER8
systems and is installed when sam2bam is built.

The accelerator plug-in  enables the use of a hardware compression card. Part of the multi-threads for compression offload compression tasks to the hardware card instead of performing compression with software. 
This accelerator plug-in is also built when sam2bam is built.

\subsubsection*{Picard Tools}

Two Picard tools, SortSam and MarkDuplicates, have been suggested to mark duplicate alignments on GATK Best Practices \cite{gatkbest-markdup}. SortSam first takes a SAM file as  input, sorts the alignments, and writes the result to a BAM file.  MarkDuplicates then takes the produced BAM file as the input, marks duplicate alignments, and writes the result to another BAM file. The BAM files are compressed by default.

We used Picard tools (version 2.1.1) \cite{picard} in the BioBuilds package (version 2016-04) \cite{biobuilds}.
OpenJDK 1.8.0\_72-internal was used to run the Picard tools with 21 GB of Java heap memory.

\subsubsection*{Hardware}
The runtime of the target programs and the maximum size of the memory that was used by the programs during  program execution were measured by using a command, {\tt /usr/bin/time}. The programs were run on a single node of IBM Power Systems S822LC
\cite{S822LC} that had 16 POWER8-based CPU cores \cite{power8}, where 128 logical processors were available (eight logical processors per core) with 1 TB of memory. The machine was attached to  high performance storage, i.e., 
IBM Elastic Storage Server (ESS) GL4 via Mellanox FDR switch \cite{ESS}.
A hardware card that provided  FPGA-based zlib acceleration \cite{GenWQE} was attached to the machine and this could speed up compression of the BAM data. The operating system was Ubuntu 14.04.1. 

Theoretical maximum performance for the storage mode of sam2bam explained in \nameref{virtualizing} was measured by using the file system in the main memory ({\tt /dev/shm}), which simulated an ideal high-performance device (e.g., a solid state drive (SSD)). Such a device is mandatory to achieve high levels of performance in the storage mode since sam2bam in the storage mode performs a huge number of I/O operations that are not always sequential accesses.

\subsection*{Performance Evaluation with a Whole Exome Data}

The sam2bam demonstrated more than 100 times better performance than Picard and finished in about one minute in both memory and storage modes while Picard needed more than two hours (Table \ref{WEX_result}).

\begin{table}[h]
\begin{adjustwidth}{-2.25in}{0in}
\caption{
{\bf Runtimes and  maximum memory sizes for marking duplicates on  52 GB WEX data.} }
\begin{tabular}
{l|p{3.9cm}|p{3.3cm}||r}
\hline
                  & {\bf SAM parsing, sorting} & {\bf Duplicate marking} & {\bf Total runtime}\\ \hline
Picard     &  59.4 min (17.5 GB) & 86.6 min  (22.2 GB) & {\bf 146.0 min} \\ \hline 
Sam2bam (memory mode) & \multicolumn{2}{|l||}{1.0 min (130.3 GB)} & {\bf 1.0 min}  \\ \hline 	
Sam2bam (storage mode) & \multicolumn{2}{|l||}{1.0 min (105.4 GB)} & {\bf 1.0 min}  \\ \hline 
Sam2bam (memory mode, no HW compression) & \multicolumn{2}{|l||}{1.4 min (129.0 GB)} & {\bf 1.4 min}  \\ \hline 
Sam2bam (storage mode, no HW compression) & \multicolumn{2}{|l||}{1.3 min (103.9 GB)} & {\bf 1.3 min}  \\ \hline 
\end{tabular}
\begin{flushleft} 
Although sam2bam was more than 
186 times 							
faster than the standard tools, it required more memory than Picard in the memory mode. The sam2bam reduced the maximum memory size by placing 
 binary alignments in external storage instead of in the main memory in the storage mode. 
 The performance of sam2bam with  data compression by both software and hardware was 43\% better than that of sam2bam with  software-only compression. 
 The Java heap size  for Picard was sufficient since the time spent in garbage collection of the Java heap was negligible
  (about 0.63\% of the total runtime).   		
\end{flushleft}
\label{WEX_result}
\end{adjustwidth}
\end{table}

The sam2bam benefited from multi-threading and pipelining, which was explained in \nameref{sam_parsing}. 
The sam2bam read a SAM file for SAM parsing and parsed it at rates of 1.7-2.0 GB/s.
This high level of performance was due to multi-threading and pipelining (we will discuss  performance without them in \nameref{S1_Text}).
The runtime of duplicate marking was ~5\% of the total runtime. 

BAM blocks were compressed at rates of 1.5-1.7 GB/s, including a rate of additional 0.9 GB/s with  hardware compression.
Such a high throughput was achieved by pipelining 
as well as multi-threading (we will discuss  performance without pipelining in \nameref{S2_Text}).

The performance of the framework on which alignments were analyzed and processed is critical for high performance tools. The runtime of a Picard tool that converts the file format from SAM to BAM is 92\% that of SortSam and 70\% that of MarkDuplicates for Picard (we will discuss the details in \nameref{S3_Text}).

\subsection*{Performance Evaluation with  Whole Genome Sequencing Data}

The sam2bam demonstrated
156 times 							
 better performance than Picard. The sam2bam finished in about ~9 minutes in the memory mode while Picard needed more than 20 hours (Table \ref{WGS_result}).

The storage mode was 81\% slower than the memory mode for  WGS data, while the memory  and storage modes demonstrated similar performance for  WEX data. This slowdown was mainly due to slowdown in  BAM block compression in the storage mode (37\% of  throughput in the memory mode). We collected the system-level profiles to analyze the slowdown in  BAM block compression. The profiles indicated that the computation time in the operating system was significantly increased by 33 times in the storage mode using the WGS data,
 but it was only  increased by  151\% when using the WEX data. We need to further investigate  additional activities undertaken by the operating system to address the slowdown with the WGS data in the storage mode.

\begin{table}[h]
\begin{adjustwidth}{-2.25in}{0in}
\caption{
{\bf  Runtimes and  maximum memory sizes for marking duplicates on  546GB WGS data.}}
\begin{tabular}
{l|p{3.9cm}|p{3.3cm}||r}
\hline
                  & {\bf SAM parsing, sorting} & {\bf Duplicate marking} & {\bf Total runtime}\\ \hline
Picard      &  631.9 min (18.6 GB) & 707.5 min  (22.4 GB) & {\bf 1339.4 min} \\ \hline 
Sam2bam (memory mode) & \multicolumn{2}{|l||}{8.6 min (710.7 GB)} & {\bf 8.6 min} \\ \hline 	
Sam2bam (storage mode) & \multicolumn{2}{|l||}{15.6 min (232.9 GB)} & {\bf 15.6 min} \\ \hline 
Sam2bam (memory mode, no HW compression) & \multicolumn{2}{|l||}{16.2 min (709.1 GB)} & {\bf 16.2 min} \\ \hline 	
Sam2bam (storage mode, no HW compression) & \multicolumn{2}{|l||}{21.7 min (231.3 GB)} & {\bf 21.7 min} \\ \hline 	
\end{tabular}
\begin{flushleft}  \end{flushleft}
\label{WGS_result}
\end{adjustwidth}
\end{table}

\subsection*{Accuracy of Duplicate Marking}
The accuracy of duplicate marking for sam2bam could be evaluated by measuring the number of  alignments that Picard MarkDuplicates marked but sam2bam did not and also by measuring the number of the alignments that sam2bam marked but PicardMarkDuplicates did  not. Outputs were compared between Picard MarkDuplicates and sam2bam (we will discuss how we compared the outputs in \nameref{S4_Text}) to evaluate the accuracy of duplicate marking. If sam2bam and Picard MarkDuplicates marked the same sets of  alignments, sam2bam could be considered to be accurate and could be used as a fast alternative to Picard MarkDuplicates.

We tested and verified that duplicate marking by sam2bam was {\it accurate}, based on the experimental results obtained from WEX and WGS data sets.
There were ~16 million duplicate alignments for the  WEX data set and ~188 million for the WGS data set. These alignments were the same between sam2bam and Picard MarkDuplicates when making the comparison explained in \nameref{S4_Text}.

\nolinenumbers

\section*{Supporting Information}

\subsection*{S1 Text.}\label{S1_Text}{\bf Evaluation of  performance of  multi-threaded and pipelined SAM parsing.}
\ifthenelse{\value{submission} > 0}{ }{

\ifthenelse{\value{submission} > 0}{
\section*{S1 Text - Evaluation of  performance of  multi-threaded and pipelined SAM parsing}
}{ }

We evaluated the performance of  sequential SAM parsing by measuring the runtime of only the SAM parsing part in samtools  to evaluate the advantage of multi-threaded and pipelined code for SAM parsing. The code modules used for receiving the input data in the SAM format and producing the binary alignments were the same for samtools and sam2bam. However, samtools called the code modules in a sequential loop while sam2bam called them in multi-threaded steps of the pipeline. We could evaluate the performance of  sequential SAM parsing by using a modified version of samtools that did not include the steps after SAM parsing.

The resulting runtimes indicated that  sequential SAM parsing only provided 8-9\% of the SAM parsing rate of sam2bam. This comparison demonstrated that sam2bam benefited from multi-threading and pipelining for SAM parsing.

}

\subsection*{S2 Text.}\label{S2_Text}{\bf Evaluation of  performance of  multi-threaded and pipelined generation of  compressed BAM file.}
\ifthenelse{\value{submission} > 0}{ }{

\ifthenelse{\value{submission} > 0}{
\section*{S2 Text - Evaluation of  performance of  multi-threaded and pipelined generation of  compressed BAM file}
}{ }

We evaluated the performance of not-pipelined code by measuring the runtime of generating a compressed BAM file in samtools to evaluate the advantage of the pipelined code that generates a compressed BAM file. An uncompressed BAM file was used to remove the cost of SAM parsing. The standard library that was installed on the operating system ({\tt zlib}) was used to compress BAM blocks on both samtools and sam2bam. 

The steps in generating a compressed BAM file are not pipelined in samtools, which has a main loop that sequentially performs  following steps: grouping the binary alignments into blocks until a buffer is full, compressing the blocks on the buffer, and writing the compressed blocks to a file.  Compressing the blocks is the most time-consuming step, but it can be done by using multi-threads.

The experimental results revealed that the throughput of samtools for generating a compressed BAM file was 38\% that of sam2bam without the hardware compression. 
This is because multi-threads for compression can only run  periodically when the buffer is full. The steps in generating a compressed BAM file to continuously run the compression threads should be pipelined so that the threads can compress the blocks without pauses. That is, the compression threads should receive the blocks from a separate thread that groups the alignments into blocks. Also, when they finish  compression, they should receive the next blocks without having to wait until the current compressed blocks have been written to a file.

}

\subsection*{S3 Text.}\label{S3_Text}{\bf Discussion on performance of  file format conversion framework.}
\ifthenelse{\value{submission} > 0}{ }{

\ifthenelse{\value{submission} > 0}{
\section*{S3 Text - Discussion on performance of  file format conversion framework}
}{ }

The performance of  tool frameworks is critical for the total performance of  analysis and processing of alignments on the tools. The tool frameworks are the software components that read  input files, provide  data to the tool cores, receive the results from the tool cores, and write the results to  output files. 

Table \ref{sam2bam} summarizes the runtimes of converting the file formats of the whole exome data from SAM to compressed BAM on Picard and samtools. Compared with the total runtimes of the Picard tools of SortSam and MarkDuplicates 
(59.4 minutes for the former and 86.6 minutes for the latter), 
which perform some work in addition to the framework's format conversion, the runtime of the Picard framework
(55.2 minutes) 
is a large part of the total runtime.

The samtools' framework performs 
57\% better 					
than the Picard's framework on the single-thread configuration. In addition, the performance of the samtools' framework can be improved by accelerating generation of a compressed BAM file by using multi-threads. However, the  improvement in performance with multi-threads does not scale well. Sixteen threads on 16 cores only achieved 3.5x speed-up.

\begin{table}[h]	
\caption{
{\bf Runtimes of  file format conversion from SAM to compressed BAM for  whole exome data.}}\label{sam2bam}
\begin{tabular}{l|l|r|r}
\hline
{\bf Tool}                                  & {\bf \#Threads} & {\bf Runtime} & {\bf Relative runtime} \\ \hline
Picard SamFormatConverter	& 1 	& 55.2 min \\ \hline 
Samtools view			& 1	& 35.2 min & 1.00 \\ \hline     
Samtools view			& 2	& 20.6 min & 0.59 \\ \hline 	
Samtools view			& 4	& 13.6 min & 0.39 \\ \hline 	
Samtools view			& 8	& 10.5 min & 0.30 \\ \hline 	
Samtools view			& 16	& 10.0 min & 0.28 \\ \hline 	

\end{tabular}
\end{table}

}

\subsection*{S4 Text.}\label{S4_Text}{\bf Methodology of comparing  marked duplicates between  tools.}
\ifthenelse{\value{submission} > 0}{ }{

\ifthenelse{\value{submission} > 0}{
\section*{S4 Text - Methodology of comparing  marked duplicates between  tools}
}{ }

The duplicate marking step marks the alignments of duplicate reads by setting a specific bit of the SAM format to one for the alignments (bit 0x400 of the FLAG field \cite{sambam}). 
We first extract the alignments whose duplicate bits are one from the output of each tool to compare the outputs between two tools for duplicate marking.

The next step is to normalize the alignments, which involves three sub-steps: sorting  optional fields of the SAM format \cite{sambam} (e.g., in  alphabetical order), removing the optional fields that were added by the tools and are not included in either output, and removing the version of the SAM format. 

The last step is file comparison. For example, the {\tt diff} command is available for  file comparison in the Linux environment. When sam2bam does not emulate the sorting of  Picard SortSam, the order of the marked alignments in the SAM file is also normalized before  file comparison so that the same alignment appears in the same position in the files. For example, the {\tt sort} command is available for this normalization in the Linux environment.

The content of the output files produced by the Picard tools and sam2bam included the minor differences that were previously explained, even where the same alignments were  marked as duplicates. The Picard MarkDuplicate tool changes the order of the optional fields, while sam2bam retains the original order since it uses the utility functions of samtools, which does not change the order of the optional fields.
Also, the Picard MarkDuplicate tool adds an optional field to indicate that MarkDuplicates has processed the file, while sam2bam does not add a MarkDuplicates-specific field. In addition, the Picard MarkDuplicate tool updates the version number in the header of the output, while sam2bam keeps it.

The duplicate marking tools that use the algorithm \cite{bamUtil} can find the same set of alignments that include the best alignment and duplicate alignments. The best alignment has the highest base quality while duplicate alignments have the base qualities that are lower than the highest base quality. 
More than a few alignments have the same highest base quality in some cases. Any such alignments can be the best alignment since its base quality is the highest in the alignment set.
However, sam2bam chooses the same best alignment from multiple candidates as Picard MarkDuplicates does by emulating how Picard MarkDuplicates chooses it from an input file that Picard SortSam has sorted. Therefore, the data comparison that was previously explained  produces no differences if sam2bam marks appropriate alignments. 

\ifthenelse{\value{submission} > 0}{

}{ }
}

\end{document}